\documentclass[universe,article,accept,pdftex,moreauthors]{Definitions/mdpi} 

\firstpage{1} 
\pubvolume{1}
\issuenum{1}
\articlenumber{0}
\pubyear{2025}
\copyrightyear{2025}
\datereceived{ } 
\daterevised{ } 
\dateaccepted{ } 
\datepublished{ } 
\hreflink{https://doi.org/} 

\def\aa{Astron. Astrophys.}

\def\apj{Astrophys. J.}

\def\mnras{Mon. Not. R. Astron. Soc.}
\def\pasj{Publ. Astron. Soc. Jap.}

\def\sp{Solar Phys.}

\Title{
{Some} 
 Questions and Challenges in Measurements of Solar Magnetic Fields}

\TitleCitation{Some Questions and Challenges in Measurements of Solar Magnetic Fields}

\Author{
{Hongqi Zhang} 
 $^{1,}$*, Jiangtao Su $^{1,2}$,  Xiaofan Wang $^{1}$, 
 Xingming Bao 
 $^{1}$, Yu Liu $^{3}$, Yingzi Sun $^1$ and Mingyu Zhao $^4$}

\AuthorNames{Hongqi Zhang, Jiangtao Su, Xiaofan Wang, 
Xiangming Bao;
 Yu Liu, Yingzi Sun and Mingyu Zhao}

\AuthorCitation{
{Zhang H.;} 
 Su J.; Wang X.; Bao, X.; Liu Y.; Sun Y.; Zhao M.,}

\address{%
$^{1}$ \quad National Astronomical Observatories of Chinese Academy of Sciences, Beijing 100101, China; 
\mbox{sjt@bao.ac.cn (J.S.); wxf@bao.ac.cn (X.W.); xbao@bao.ac.cn (X.B.); syz@bao.ac.cn (Y.S.)} 
 \\
$^{2}$ \quad University of Chinese Academy of Sciences, Beijing 100049, China \\
$^{3}$ \quad School of Physical Science and Technology, Southwest Jiaotong University, Chengdu 610031, China; 
lyu@swjtu.edu.cn\\
$^{4}$ \quad  Yunnan Observatories,  Chinese Academy of Sciences, Kunming 650215, China; 
myzhao@ynao.ac.cn}

\corres{Correspondence: hzhang@bao.ac.cn}

\abstract{Huairou Solar Observing Station of the National Astronomical Observatories of Chinese Academy of Sciences has been in operation since 1987. During its operation, successful observations of the solar vector magnetic field have been conducted. On the basis of the achievements at Huairou, we analyze the methods of observing the solar magnetic field, including discussions of the approximation of the transfer theory of the solar magnetic field in the atmosphere, wide field of view polarized observation, and some questions on the inversion of solar magnetic field data. We also present relevant challenges for further research.}

\begin{document}

\section{Introduction}

The American astronomer \citet{Hale08} first discovered the existence of the magnetic field in sunspots by using the Zeeman effect in 1908. In the 1950s, \citet{babc53,Babcock55} observed that there was a universal magnetic field on the solar surface of the Sun by using a photo-electromagnetic imager. In the 1960s, Severny et~al. \cite{SeBu8,Sev62} of the Crimean Observatory in the former Soviet Union carried out the first vector magnetic field measurements and observations in the solar active region. The Japanese solar physicist Unno \cite{Unno56} and Rachkovsky \cite{Rachkovsky62b} of the former Soviet Union took the lead in theoretically analyzing the radiation transfer of the polarized spectrum in the solar atmosphere during the 1960s. These studies laid the foundations for the measurement and theory of solar magnetic fields. 

The energy source and trigger mechanism of solar flares, coronal mass ejections, and other violent activities are closely related to the solar magnetic field. Observations of the solar magnetic field are helpful in revealing the process of accumulation, release, and transmission of magnetic energy during these activities. For example, by observing the changes of magnetic field it is possible to understand how the magnetic field distorts and accumulates energy before the flare explosion as well as how the energy is released quickly during the~explosion. 

The scientific observation of solar magnetic fields in China began in the 1980s with the Solar Magnetic Field Telescope led by \citet{Ai89}. This telescope can observe the solar vector magnetic field, chromosphere magnetic field, and velocity fields. A series of important research results have been obtained, including Huairou Solar Observing Station of the National Astronomical Observatories and Big Bear Solar Observatory achieving the first continuous observation of the Sun \cite{Wangh89}. 
The successful development of these instruments has allowed China to step to the forefront of physical research on solar magnetic~fields. 

On this basis, Ai \cite{aig93} analyzed the historical developments of solar magnetic field observation around the 1990s, including Stokes spectrum observation with Zeeman effect scanning point-by-point on the Sun, rapid imaging magnetic field observation with a single-wavelength video magnetograph (the wavelength position of a single Zeeman spectrum), and the dialectical relationship of the inevitable development law of two-dimensional Stokes spectrum observation. Ai's view is of great guiding significance in the observation and research on solar magnetic fields. 

At the beginning of this century, the full solar disk vector magnetic field telescope (solar magnetic activity monitoring system) began successful operation at Huairou Solar Observing Station of the National Astronomical Observatories and obtained a series of observation data on the full solar disk vector magnetic field \citep{Zhang07}. Since then, observation and research on the local solar magnetic field in China has been extended to the whole solar surface, providing a strong guarantee for solar physics research and solar activity monitoring in China. 

In recent years, observation of the solar magnetic field has gradually developed from qualitative observation to analytical research, such as extrapolating the magnetic field configuration in the upper atmosphere of the Sun, for example by observing the photospheric magnetic field
 \citep{Hag77,liu11}, calculating the solar photospheric current, magnetic (current) helicity, etc., to study the development of non-potential magnetic energy in individual solar activity regions \citep{Bao00,Yangx12}, and investigating the internal relationship between the magnetic helicity in the photosphere and the internal dynamo process of the Sun \citep{ketal03,Zhang10,2012ApJ...751...47Z}. 

In-depth study of the solar magnetic field always results in some differences between the results of different observation instruments \citep[][]{Wangh89,Bao00,Pevtsov05,Xu16,Xu21}. This difference in observations often leads to different explanations, and can cause confusion. The source of these observation differences is essentially an important topic that attracts attention to the accuracy of measurements of the solar magnetic field, especially measurements of the full-solar disk vector magnetic field with wide field of view optical systems, such as those by \citet{Wangx10}. Below, we discuss the sources of these differences and measurement accuracy more generally, hoping to benefit future studies of the solar magnetic field.

\section{Approximation Theory in  Measurement of Magnetic Fields}

 Today, observation of the solar magnetic field is usually concentrated in the lower atmosphere of the Sun, and the observed solar magnetic field is analyzed using the polarization spectrum produced by the Zeeman effect of spectral lines in the magnetic field. The general method is to use the four Stokes eccentric vibration parameters $I$, $Q$, $U$, and $V$ for analysis, defined as the Stokes spectrum. It should be pointed out that the solar spectrum reflects more information about the physical state of the solar atmosphere, and that the information about the magnetic field in the solar atmosphere is mixed in the spectrum. 
 In other words, the challenge is how to obtain the message of the magnetic field from the spectrum in the solar atmosphere. In the process of diagnosing the solar magnetic field, in theory the first key problem should be how to analyze and simplify this connection.  
 
It should be pointed out that the first question encountered in magnetic field measurement involves the formation of spectral lines in the solar magnetic field atmosphere~\cite{Kawakami83}. The theory of polarized light radiation transfer can be said to be an approximate theory. Here, the level of theoretical approximation is key, including whether to consider light scattering, the local thermal dynamic equilibrium (LTE) approximation, how to determine the approximation of temperature and particle density in the solar atmospheric model, and the possible uniformity or distribution of the Sun's magnetic field. Usually, a relatively simplified form of the radiative transfer equation is used for calculating and analyzing the polarized light in the solar magnetic field atmosphere. These solutions are naturally approximate, and are included in the theoretical analysis of radiative transfer in solar model atmosphere \citep{Ai82,ZhangA87,SongW90,SongW92,SongW93,Zhang19,Zhang20} and observing data reduce process \citep{WangTJ96,Su04b,Su05,Wangx10} employed at Huairou Solar Observing Station.

In the solar atmosphere, the non-LTE equation for the transfer of polarized radiation {can be written in matrix form (\mbox{
eq. (6.54) of}})~\citep{Stenflo94}:
\begin{equation}
\label{eq:transij}
\frac{d{\bf I}_\nu}{dt}+\frac{d{\bf I}_\nu}{ds}=-(N_{J_l}B_{J_lJ_u}\Phi_{J_l}-N_{J_u}B_{J_uJ_l}\Phi_{J_u})\frac{h\nu}{4\pi}{\bf I}_\nu+{\bf j}
\end{equation}
{
where 
 ${d{\bf I}_\nu}/{dt}$ has been introduced.} In the case of a non-scattering process, the emission vector can be approximately by
\begin{equation}
\label{eq:trasj}
{\bf j}=\frac{h\nu}{4\pi}\Phi_{J_u}{\bf 1}N_{J_u}A_{J_uJ_l},  
\end{equation}
where $\bf I$ represents the Stokes parameters ($I$, $Q$, $U$, $V$), {\color{blue} $\bf 1$ is the unit four-vector, $\bf J$ is the emission vector,} $N_{J_u}$ and $N_{J_l}$ are the respective densities of the upper-level and lower-level particles, $\Phi_{J_u} $ is the absorptive coefficient matrix, and $A_{J_uJ_l}$, $ B_{J_uJ_l}$, and $ B_{J_lJ_u}$ are Einstein's coefficients. Equation (\ref{eq:transij}) only describes the absorption and emission processes in the spectral line. 

The complete steady state radiation transfer equation {($d{\bf I}_\nu/{dt}=0$) should include the contributions of the line spectrum and continuous spectrum.  It can be assumed that the existence of the magnetic field does not change the frequency dependence of the absorptive coefficient and emission coefficient of the continuum. Therefore, the contributions of the two parts are added together. For electric--dipole transitions, assuming that the spectral line originates in the transition between two levels of an atom and no atomic polarization is present in the two levels, the  radiation transfer equation can be written in the following form \citep{Land04}:     
\begin{equation}
\label{eq:transfer}
\begin{aligned}
\frac{d}{ds}\left(\begin{array}{c}
I\\
Q\\
U\\
V
\end{array}\right)=&-k_c\left(\begin{array}{cccc}
1&0&0&0\\
0&1&0&0\\
0&0&1&0\\
0&0&0&1
\end{array}\right)\left(\begin{array}{c}
I-S_C\\
Q\\
U\\
V
\end{array}\right)\\   
&-k_L\left(\begin{array}{cccc}
\phi_I&\phi_Q&\phi_U&\phi_V\\
\phi_Q&\phi_I&\psi_V&-\psi_U\\
\phi_U&-\psi_V&\phi_I&\psi_Q\\
\phi_V&\psi_U&-\psi_Q&\phi_I
\end{array}\right)\left(\begin{array}{c}
I-S_L\\
Q\\
U\\
V
\end{array}\right)
\end{aligned}
\end{equation}
where $k_c $ is the continuous spectral absorption coefficient, $k_L$ is the spectral line absorption coefficient (corrected for stimulated emission), $\phi_{(I,Q,U,V)}$ are the absorptive parameters, and $\psi_{(Q,U,V)}$ are the anomalous dispersion Stokes parameters $(I,Q,U,V)$. Here, $S_c$ is the source function of the continuous spectrum, which in most cases is equivalent to the Planck function, while $S_L$ is the source function of the spectral line.

When studying radiation transfer in the solar magnetic field atmosphere, it is usually necessary to simplify the radiation transition process of atoms in the magnetic field, such as by simplifying the interference effect between the magnetic energy sublevels, assuming that these levels are in a statistical equilibrium state, etc. \cite{Stenflo94,Land04}. Often, the micro and macro states in the solar atmosphere are simplified to varying degrees, for example in \citep{Stix02,Zhang19, Zhang20}. The research work in this area still needs to be further deepened.

An important analytical solution of the radiation transfer equation for polarized radiation was provided by Unno \citep{Unno56}. This solution contains several limitations: 
(a) a semi-infinite dimensionless plane parallel atmosphere with a constant magnetic field; (b) LTE, and the Planck function is linear in the continuum optical depth measured along the vertical;
 (c) a constant ratio of linear and continuous absorption coefficients; and (d) the propagation matrix is depth-independent. Under these assumptions, the polarization radiation transfer equation can be solved as follows \citep{Stenflo94}:       
\vspace{-9pt}
\begin{adjustwidth}{-\extralength}{0cm}
\begin{eqnarray}
\label{eq:absop}     
&~&I=B_0+\mu B_1\Delta^{-1}[(1+\eta_I)((1+\eta_I)^2+\rho_Q^2+\rho_U^2+\rho_V^2)],\nonumber\\
&~&Q=-\mu B_1\Delta^{-1}[(1+\eta_I)^2\eta_Q+(1+\eta_I)(\eta_V\rho_U-\eta_U\rho_V)+\rho_Q(\eta_Q\varrho_Q+\eta_U\varrho_U+\eta_V\varrho_V)],\nonumber\\
&~&U=-\mu B_1\Delta^{-1}[(1+\eta_I)^2\eta_U+(1+\eta_I)(\eta_Q\rho_V-\eta_V\rho_Q)+\rho_Q(\eta_Q\varrho_Q+\eta_U\varrho_U+\eta_V\varrho_V)],\nonumber\\
&~&V=-\mu B_1\Delta^{-1}[(1+\eta_I)^2\eta_V+(1+\eta_I)(\eta_Q\rho_V-\eta_V\rho_Q)+\rho_V(\eta_Q\varrho_Q+\eta_U\varrho_U+\eta_V\varrho_V)],
\end{eqnarray}
\end{adjustwidth}
where
\vspace{-9pt}
\begin{adjustwidth}{-\extralength}{0cm}
\begin{eqnarray*}
&~&\Delta=(1+\eta_I)^2[(1+\eta_I)^2-\eta_Q^2-\eta_U^2-\eta_V^2+\rho_Q^2+\rho_U^2+\rho_V^2)]-(\eta_Q\varrho_Q+\eta_U\varrho_U+\eta_V\varrho_V)^2
\end{eqnarray*}
\end{adjustwidth}
in which $B(\tau_c)=B_0+B_1\tau_c$, $\mu$ is the cosine of the angle between the ray path and the outward direction of the atmosphere,  $\eta_{(I,Q,U,V)}$ are the absorptive coefficients, and $\rho_{(Q,U,V)}$ are the Stokes anomalous dispersion coefficients $(I,Q,U,V)$.

\textls[-25]{In the approximation of the weak magnetic field \citep{Stix02}, it is possible to obtain the~formulas}    
\begin{equation}
\label{eq:weakapprox}
\begin{aligned}
\eta_I&=\eta_p+O(v_b^2),\\ 
\eta_Q&= - \frac{1}{4}\frac{\partial^2\eta_p}{\partial v^2}v_b^2\sin^2\psi \cos2\varphi +O(v^4_b) ,\\
\eta_U&=- \frac{1}{4}\frac{\partial^2\eta_p}{\partial v^2}v_b^2\sin^2\psi \sin2\varphi +O(v^4_b) ,\\
\eta_V&=\frac{\partial\eta_p}{\partial v}v_b\cos\psi +O(v^3_b),
\end{aligned} 
\end{equation}
where $\psi$ and $\varphi$  are the magnetic inclined angle and azimuth angle, respectively, and the magnetic displacement is
$v_b=\Delta\lambda_B/\!\Delta\!\lambda_D$. 

Using Equation~(\ref{eq:absop}) and neglecting the magneto-optic effects, we can obtain the simple relationship between the magnetic field and the Stokes parameters. It is found that
\begin{equation}
\label{eq:weakapproxa}
\begin{aligned}  
I\approx&B_0+\frac{\mu B_1}{(1+\eta_I)},\\ 
Q\approx&-\frac{\mu B_1}{(1+\eta_I)^2}\eta_Q\approx C_T'B^2\sin^2\psi \cos2\varphi =C_T'B^2_T \cos2\varphi,\\
U\approx&-\frac{\mu B_1}{(1+\eta_I)^2}\eta_U\approx C_T'B^2\sin^2\psi \sin2\varphi  =C_T'B^2_T \sin2\varphi ,\\
V\approx&-\frac{\mu B_1}{(1+\eta_I)^2}\eta_V\approx C_L'B\cos\psi =C_L'B_L, 
\end{aligned}  
\end{equation}
where 
\begin{equation}
\begin{aligned}  
\label{eq:weakapproxb}
C_T'    &=  \frac{\mu B_1}{4(1+\eta_I)^2}\frac{\partial^2 \eta_p}{\partial v^2} \left[\frac{e\lambda_0^2}{4\pi m_ec^2\!\Delta\!\lambda_D}(M_1g_1-M_2g_2)\right]^2,\\
C_L'    &=- \frac{\mu B_1}{(1+\eta_I)^2}\frac{\partial \eta_p}{\partial v} \frac{e\lambda_0^2}{4\pi m_ec^2\!\Delta\!\lambda_D}(M_1g_1-M_2g_2).   
\end{aligned}   
\end{equation} 

{\color{red}This} 
 means that the longitudinal component of the magnetic field can be written in the following form:
\begin{equation}
B_L=C_L V\label{eq:weekfild1} 
\end{equation}
where $C_L$ is the calibration parameter for the longitudinal component of the field, which is a function of the spectral line wavelength. This allows us to obtain the strength $B_T$ and azimuth angle $\varphi$ of the transverse magnetic field:
\begin{equation}
\label{eq:weekfild2}
B_T=C_T \sqrt[4]{Q^2+U^2},\\~~~~~~~~~~
\varphi=\frac{1}{2}\tan^{-1}\left(\frac{U}{Q}\right).
\end{equation}

It can be seen that Equation (\ref{eq:absop}) along with (\ref{eq:weekfild1}) and (\ref{eq:weekfild2}) provide theoretical solutions for polarized light in the solar magnetic field atmosphere at different degrees of approximation. These solutions are commonly used in the inversion process of magnetic field observations. This indicates that when measuring the solar magnetic field using the Stokes parameters, we often obtain an approximate value. The degree of approximation depends on the conventions in the above theories and the selection of solar atmospheric physical parameters. From Formulas (\ref{eq:weekfild1}) and (\ref{eq:weekfild2}), it can be seen that the longitudinal Stokes signal $\mathrm{V}$ and transverse magnetic field $\mathrm{Q,U}$ of the Sun are at different levels of amplitude.

It is worth noting that measuring the solar magnetic field requires complex knowledge of both the spectrum in the solar atmosphere and the observing instruments. The former relates to the influences of the solar limb darkening effect, differential rotation, velocity field on the solar surface, and thermodynamic parameters of various structures (sunspots, granulations, faculae, etc.) on measurements of the magnetic field, while the latter relates to the performance of solar observational instruments.
These complexities lead to notable question around the causes of possible differences in solar vector magnetogram observations made at different observatories \cite{ZhangHQ03}.

\section{Questions Due to Weak Field Approximation and Polarized Spectral Analysis}

When using the weak field approximation method to calibrate the solar magnetic field, it is important to note its effective range. Saturation may occur when measuring a strong magnetic field such as the umbra of sunspots, meaning that the magnetic field strengthens but the polarization signal weakens {at a certain wavelength}. Because the effective Land\'e factor $g_{eff} $ of the working spectral line is higher, this results in variation of the nonlinearity range of magnetic field measurements, such as changes in the video magnetograph at the single wavelength \cite{Zhang23}.

For example, the  FeI$\lambda {6302.5} $ spectral line $g_{eff}=2.5 $ has an effective linearity range of less than 1000 Gauss (0.06\AA {} from the line center) for longitudinal magnetic field and a range of less than 500 Gauss (0.12\AA {} from the line center) \citep{Land04} for the transverse magnetic field, while the Land\'e factor of the FeI $\lambda {5324.19} $\AA {} spectral line is $g_{eff}=1.5 $. At a deviation of 0.12\AA {} from the center of the line, its linear range for the longitudinal magnetic field is an order of 2000 Gauss, while the transverse magnetic field linear range calculated under the quiet atmosphere model is that of 1000~Gauss. The linearity range of the longitudinal magnetic field calculated by the umbra atmospheric model is less than 1000 Gauss, while that of the transverse magnetic field is about an order of 600 Gauss \citep{Zhang19}.  It is notable that the FeI $\lambda {5324.19} $ is a weak saturated line in the solar magnetic field even in the order of 3000 Gauss, as indicated by \citet{Ai82} in 1982 with theoretical analysis. FeI $\lambda {5324.19} $\AA {}  is a working line for the measurements of the photospheric magnetic fields at Huairou Solar Observing Station in China \cite{Ai89,aig81,Ai82}.

Figure \ref{fig:saturation} shows Stokes I and V as observed by the Huairou 35cm Solar Magnetic Telescope (SMFT). It can be seen that the intensity of the Stokes $V$ signal in the center of the sunspot umbra is weaker than that of the surrounding penumbral signal.  This means that the magnetic field signal in the center of the sunspot will be weaker than that of the surrounding penumbra when taking the same calibration parameter. Through research, it has been found that this difference is mainly caused by the difference in atmospheric temperature between sunspots and the quiet Sun, along with other factors. 

\begin{figure}[H]
\centerline{\includegraphics[width=33pc,angle=0.0]{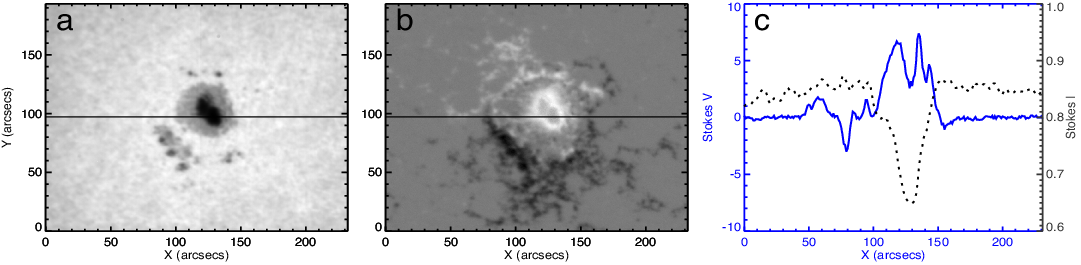}}
\caption{
{\color{red}Stokes} 
 I (\textbf{a}) and Stokes V (\textbf{b}) at FeI $\lambda {5324.19} $ \AA{}-0.075\AA{} as observed by the 35 cm diameter Solar Magnetic Field Telescope (SMFT) at Huairou Solar Observing Station, National Astronomical Observatory, Chinese Academy of Sciences. The distribution (\textbf{c}) of relative intensity values corresponds to the horizontal lines in Stokes I and V, where the solid line is Stokes V and the dotted line is the relative intensity of the Stokes I signal.
}
\label{fig:saturation}
\end{figure}

According to Formulas (\ref{eq:weakapproxa}) and (\ref{eq:weakapproxb}), we have presented the signal amplitudes of Stokes signals $V\sim dI/d\lambda$ and $(Q,U)\sim d^2I/d\lambda^2$ under the relative magnetic field intensity~\citep{Land04}, which is derived from the observed spectral profiles of the FeI$\lambda {5324.19} $\AA {} in the quiet Sun and umbra in Figure \ref{fig:obsspectrum}. The amplitudes of the Stokes signal in the umbra are weaker than in the quiet Sun with the same field strength. The ratio is about 0.3 at the peak values of the Stokes $V$ signals.
This is consistent with the observation trend in \mbox{Figure \ref{fig:saturation}c}~\citep{Plotnikov21}. In this case, similar results can also be found in the observation of the sunspot magnetic field with other filter types \citep{Chae07}. 
The possible approximation of weak fields for the measurements of transverse fields in the quiet Sun and umbra is presented in Figure \ref{fig:obsspectrum} with $(Q,U)\sim d^2I/d\lambda^2$, where the ratio is about 0.17 at the peak values of the Stokes $Q$ signals. This ratio obviously depends on the profiles of the spectral lines.
This may represent a noteworthy problem that needs to be carefully analyzed for magnetographs in the future.

\begin{figure}[H]
\centerline{\includegraphics[width=32.5pc,angle=0.0]{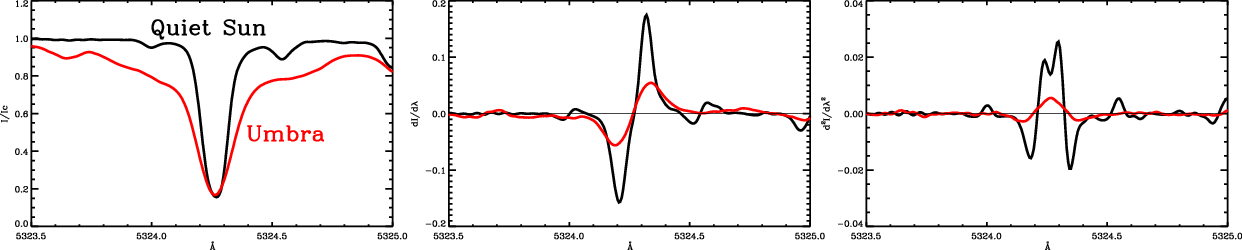}}
\caption{
(\textbf{Left}):  Observed spectral profiles of FeI$\lambda {5324.19} $\AA {} in the quiet Sun (black) and the umbra (red) from Kitt Peak Solar Observatory.
(\textbf{Middle}): The corresponding relative Stokes signals $V\sim dI/d\lambda$ inferred from the spectral line.  (\textbf{Right}): Stokes signals $(Q,U)\sim d^2I/d\lambda^2$.
}
\label{fig:obsspectrum}
\end{figure}
Subfigures a and e Figure \ref{fig:moline5324fra} respectively show the numerical calculation of Stokes $V$ in the solar model atmospheres of the quiet Sun \cite{Vernazza81} and the sunspot umbra \citep{Stellmacher70,Stellmacher75} for the FeI$\lambda$5324.19\AA{} line. It can be observed that the Stokes $V$ signal in the umbra is weaker than in the quiet Sun under the same magnetic field conditions \cite{Zhang19}. This result is consistent with the estimation of the weak field approximation presented above. This may be a noteworthy problem that needs to be carefully analyzed for magnetographs in the future, as it relates to routine~observations.

\begin{figure}[H] 
\centering 
\begin{minipage}[t]{0.49\textwidth} 
\centering 
\includegraphics[width=\textwidth]{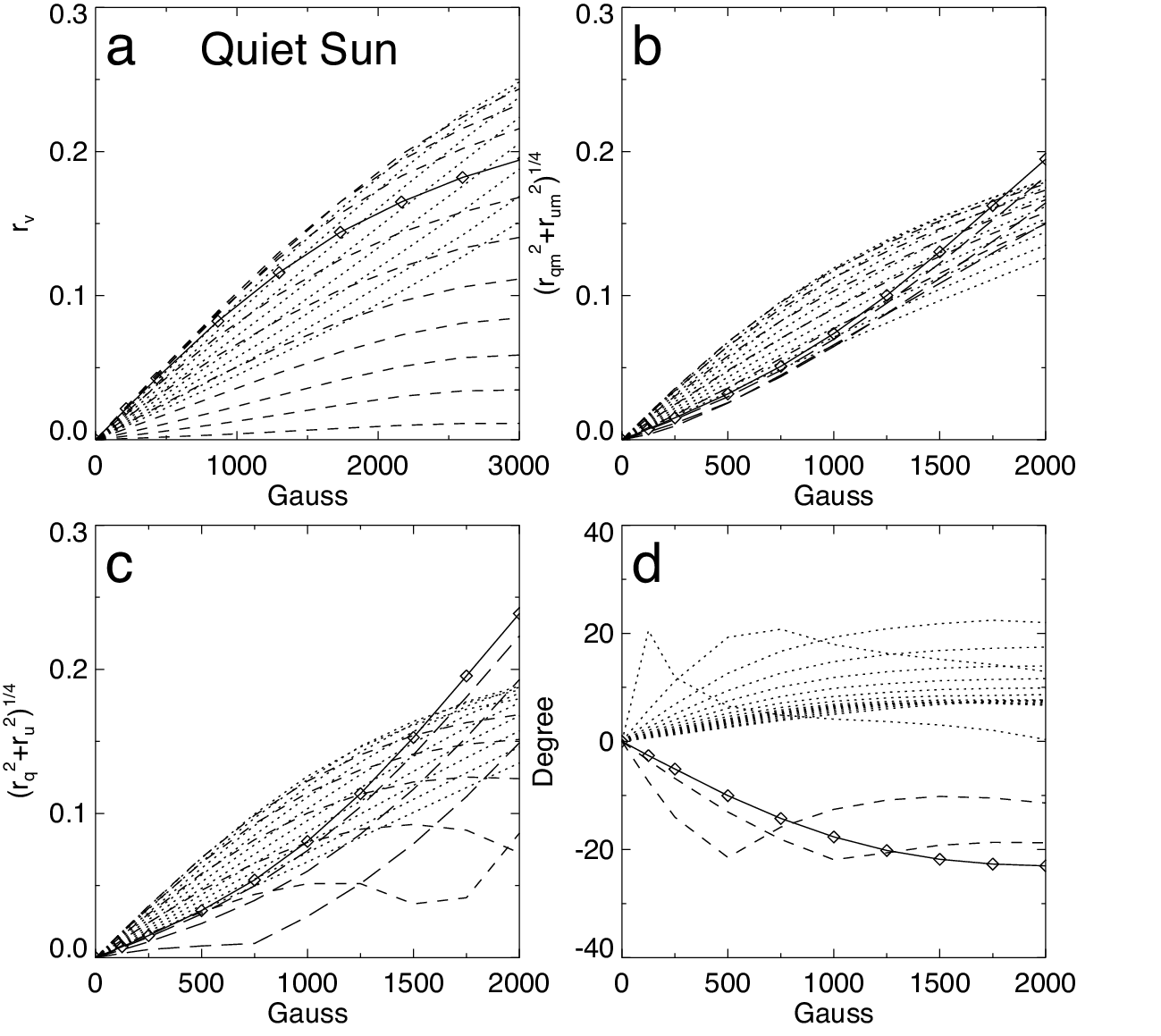} 
\end{minipage}
\hfill 
\begin{minipage}[t]{0.49\textwidth} 
\centering 
\includegraphics[width=\textwidth]{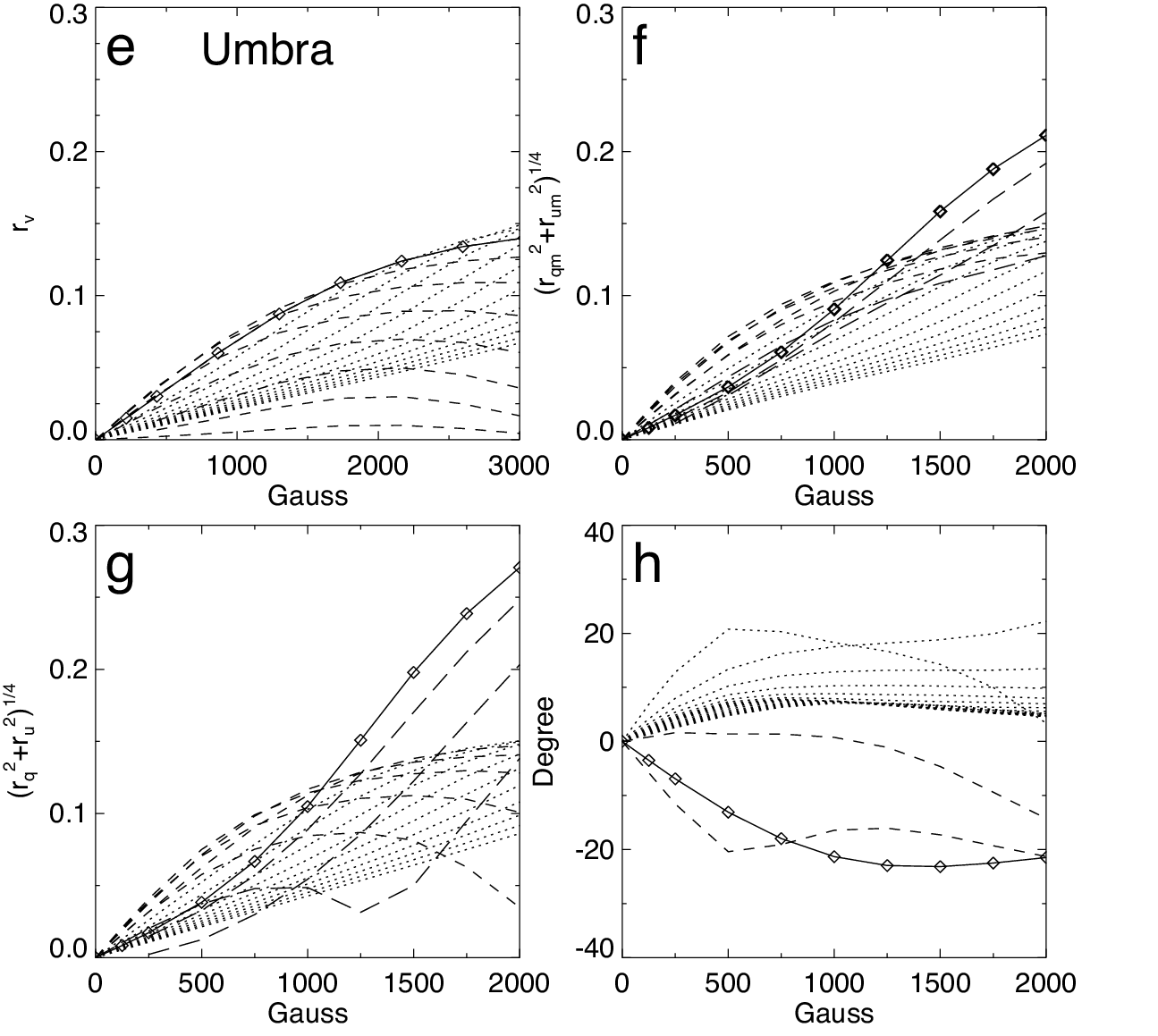} 
\label{fig:figure2} 
\end{minipage}
\caption{
{\color{red}Numerical} 
  calculation of the Stokes $V$ ($r_v$), $Q$ and $U$ $\left((r^2_q+r^2_u)^{1/4}\right)$,  and ``error azimuthal angles'' of the transverse field inferred due to the magneto-optical effect in the wing from 0.005 to 0.185\AA{} with the interval of 0.01\AA{} in the solar model atmospheres of (\textbf{a}--\textbf{d}) for 
  {the} 
 quiet Sun  \cite{Vernazza81} and {\color{red}(\textbf{e}--\textbf{h})} sunspot umbra \citep{Stellmacher70,Stellmacher75}  for the FeI$\lambda$5324.19\AA{} line with field inclination $\psi$ = 30$^\circ$, azimuth $\varphi=22.5^\circ$, and $\mu =1$.  The magneto-optical effect has been neglected in (\textbf{c}) and (\textbf{g}). {The line clusters represent the results from 0.005 to 0.185\AA{} with the interval of 0.01\AA{} from the FeI$\lambda$5324.19\AA{} line center.}  
Resource taken from \cite{Zhang19}.         
\label{fig:moline5324fra}} \end{figure}
The possibility of stray light in the instrument during magnetograph observations and the saturation effect of the receiver CCD system are also questions that need to be discussed \citep{Su05,Xu21}. 

Figure \ref{fig:Faradyr} shows the vector magnetic fields in the local active regions observed by the Full-Disk Solar Optical and Magnetic Field Monitoring System (SMAT) at the Huairou Solar Observing Station of the National Astronomical Observatory of the Chinese Academy of Sciences (first line) and by the HMI on board the SDO space satellite (second line). The data have been smoothed to the same resolution. From detailed analysis, it can be seen that the vector magnetic maps observed by the two instruments have good overall similarity, with some small differences.

Wittmann \cite{Wittmann71} pointed out that the antisymmetric elements in the second matrix on the right in Equation ($\ref{eq:transfer}$) correspond to the magneto-optical effect. The Faraday effect (induced circular birefringence) causes the electric vector of linearly polarized light to rotate by an angle which depends on the refractive index of two orthogonal modes of circularly polarized light $n$. 
The Voigt effect (induced linear birefringence, also known as Cotton--Mouton effect) leads to phase retardation between linear polarization parallel and perpendicular to the magnetic field, which depends on the corresponding refractive index difference. 
This indicates that changes in atmospheric parameters of the solar magnetic field relate to the magneto-optical effect of the spectrum \cite{Zhang23}.

It should be pointed out that the magneto-optical effect in the polarization radiative transfer Equation (\ref{eq:transfer}) is manifested as the antisymmetric term of the second propagation matrix at the right end. If the weak field limit 
is expanded to the fifth order, it can be found that the magneto-optical effect produces corrections of order 
$\left(\frac{\Delta\lambda_B}{\Delta\lambda_D}\right)^5$, 
$\left(\frac{\Delta\lambda_B}{\Delta\lambda_D}\right)^4$, and
$\left(\frac{\Delta\lambda_B}{\Delta\lambda_D}\right)^3$ 
to the Stokes parameters $I$, $V$,  and $Q-U$, respectively \cite{Land04}.

\begin{figure}[H]

\begin{adjustwidth}{-\extralength}{0cm}
\centering 
{\includegraphics[width=34pc,angle=0.0]{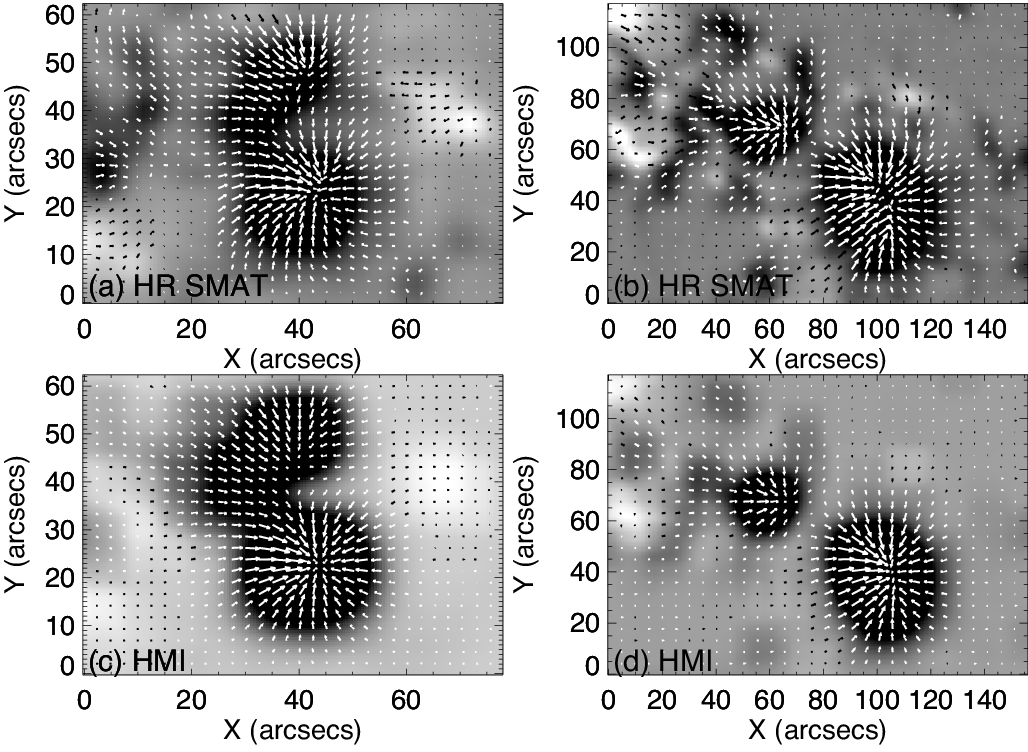}}
\end{adjustwidth}
\caption{Vector magnetograms observed by the Full-Disk Solar Optical and Magnetic Field Monitoring System (SMAT) at the Huairou Solar Observing Station of the National Astronomical Observatory of the Chinese Academy of Sciences on January 18, 2017 (first line) and by the HMI carried by the SDO space satellite (second line) on September 3, 2017. The white (black) color indicates positive (negative) polarity, while the arrows indicate the transverse components of the magnetic field.}
\label{fig:Faradyr}
\end{figure}

 The magneto-optical effect (polarization plane deflection of polarized light in magnetic media) is a problem worthy of discussion in the context of measuring magnetic fields with video magnetographs. The influence of the magneto-optical effect on magnetic field measurements has been studied in China for a long time, and has been used in the observation and analysis of magnetic field, {for example} by \citep{Ai82,Ye83}. Figure \ref{fig:moline5324fra} shows an example of the variation in the Stokes parameters $V$, $Q$, and $U$ $\left((r^2_q+r^2_u)^{1/4}\right)$ along with the ``error azimuthal angles'' of the transverse field in a model atmosphere of the quiet Sun and sunspot umbra inferred from magneto-optical effects numerically calculated using the FeI $\lambda {5324.19} $\AA {} line with the radiative transfer equations \cite{Zhang19}. Different rotation angles of the azimuthal are found at different wavelengths from the line center due to the magneto-optical effects.

\citet{Bao00} analyzed the influence of magneto-optical effect on the observation data of different solar magnetic field measuring instruments from the shape of vector magnetic field in the solar active regions. 
Figure \ref{fig:ARfit1} shows the FeI$\lambda{5324.19}$\AA{} Stokes polarization distribution obtained from the sunspot umbra observation fitted by the least square method when using Formula (\ref{eq:absop}) to analyze the Stokes polarization spectrum profile in the solar magnetic field observed by the Huairou Solar Magnetic Field Telescope. According to the theoretical calculation, it is preliminarily estimated that the deflection error angle of the transverse magnetic field for the FeI$\lambda{5324.19}$\AA{} spectral line is about 5$^\circ$--10$^\circ$, as indicated by \citet{Zhang00} and \cite{Su04a,Su04b} 
{\color{red}in} 
 most situations. These calculations provide an order of the estimations on the errors of the azimuthal angles of the transverse magnetic field from the observations of the video magnetographs. 

\begin{figure}[H]
{\includegraphics[width=32.5pc]{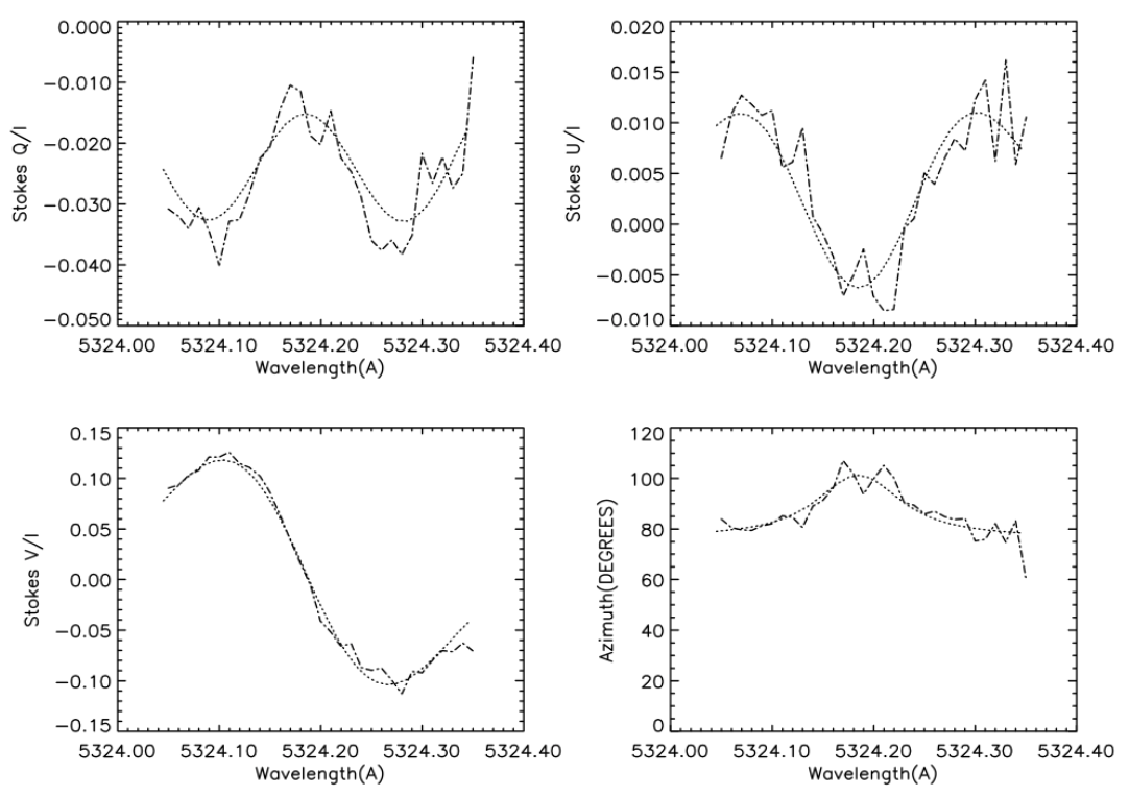}}
\caption{ 
The Stokes polarization spectral distribution of Stokes $Q/I$, $U/I$, and $V/I$ (dotted lines) for the FeI $\lambda {5324.19} $\AA {} spectral lines obtained in the umbra of sunspots using the Huairou Solar Magnetic Field Telescope compared with the fitted distribution generated by nonlinear least squares inversion (dashed lines). 
{The fourth figure shows the variation of the azimuthal angle of the transverse field 
 inferred from $Q/I$ and $U/I$ using the second formula of Equation~(\ref{eq:weekfild2}) and neglecting the magneto-optical effect \cite{Su04b}. }       
}    
\label{fig:ARfit1}
\end{figure}

\section{The Challenge of Magnetic Field Measurements in the Higher Solar~Atmosphere}

 \subsection{Chromospheric Magnetic Fields}
 
 The chromospheric magnetic field is usually considered as the result of the upward extension of the solar photospheric magnetic field, and the complexity of the structure is an interesting topic. It is generally believed that the chromospheric magnetic field plays a key role in the physical process of the chromosphere, affecting the movement and distribution of matter. Violent activities such as solar flares and eruption of the prominences are directly and closely related to the change and energy release of the chromospheric magnetic field. 

The H$\beta$ is a chromospheric line for the diagnostic of the magnetic field at Huairou Solar Observing Station~\cite{Nagaraju20}.   
A series of H$\beta$ chromospheric longitudinal magnetograms have been observed since 1987. The formation of the H$\beta$ line was analyzed by \mbox{\citet{ZhangA87,Zhang19,Zhang20}}.  As a magnetic field is present, the broadening of the hydrogen line, such as H$\beta$, should be the joint effect of the magnetic field and the microscopic electric field inclined at various angles and distributed according to the Holtsmark statistics, with the effect of nonlocal thermodynamic equilibrium (Non-LTE) in the solar atmosphere remaining a notable question \cite{Stenflo94,Land04,BaiX13,Zhang19}.  

The chromospheric magnetic field with the photospheric vector magnetic field in an active region can be found in Figure \ref{fig:ar6619}, where the reversal signal of the longitudinal component in the positive polarity of the chromospheric magnetogram is caused by the blended photospheric line in the wing of the H$\beta$. This topic has been discussed by~\citet{Zhang19,Zhang20}. Similar observational evidence of Balmer lines for the diagnostic of the chromospheric magnetic field was presented by~\citet{Balasubramaniam04,Nagaraju20}. 
 It is important to understand the distribution of the magnetic field in the higher solar atmosphere compared to the photosphere in the active regions and the relationship to solar flares.  A comparison between the chromospheric and photospheric magnetic features in the quiet Sun has been presented by \citet{Zhangm98}.  
 
\begin{figure}[H]
\includegraphics[width=125mm]{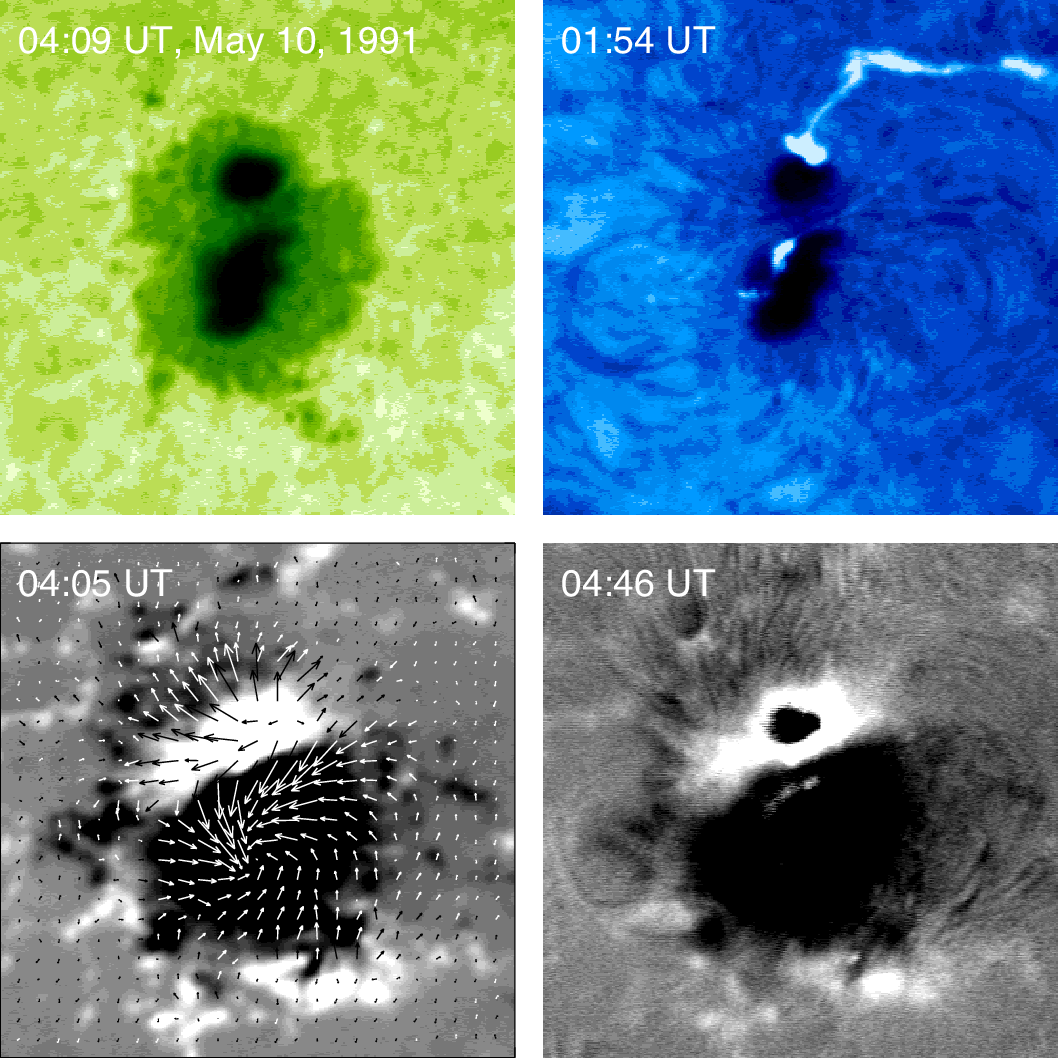}  
\caption{Active region NOAA 6619 on 10 May 1991. (\textbf{Top left}): Photospheric FeI$\lambda$ 5324.19\AA{}   filtergram. (\textbf{Bottom left}): Photospheric vector magnetogram. (\textbf{Top right}): H$\beta$ filtergram. (\textbf{Bottom right}): H$\beta$ longitudinal magnetogram. White (black) indicates positive (negative) polarity in the magnetograms. The size of each figure is $2.'8\times 2.'8$. Taken from \cite{Zhang19}.
\label{fig:ar6619}}
\end{figure}

Figure \ref{fig:contribHbeta} shows the numerical calculation on the distribution of the contribution function of the H$\beta$ line at different wavelengths from the line center under the VAL atmospheric model in the quiet Sun by \cite {Vernazza81} (left) and the sunspot umbral atmospheric model from \cite{Ding91, Zhang20}. The contribution of the Stokes $V$ signal in the H$\beta$ line in the relatively lower atmosphere in the umbra region can be seen relative to the quiet Sun with the same optical depth of the continuum \cite{Zhang19}. 
 This only provides a preliminary estimation from the numerical calculation of the atmospheric model with the magnetic fields. In the chromosphere, the direction of the magnetic field is complex and changeable. The direction of the magnetic field may differ at different positions, and there may even be a large inclined angle in the direction of the magnetic field in adjacent areas, resulting in distortion and winding of the magnetic field. In addition, the magnetic field in the chromosphere may be non-uniform and fibrous, while strong and weak magnetic field areas may be intertwined to form complex patterns. These issues make discussion of the solar chromospheric magnetic field more~complicated.
 
\begin{figure}[H]
{\includegraphics[width=90mm]{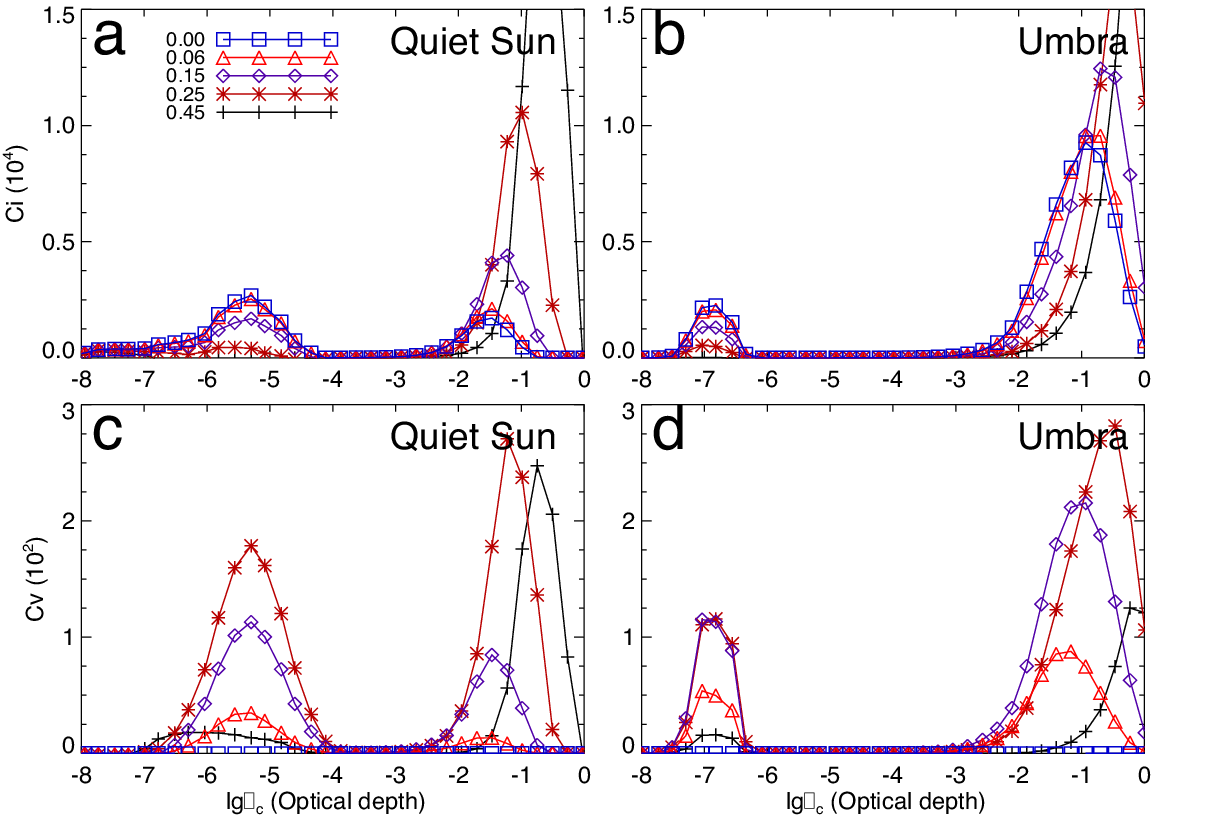} } 
\caption{
{\color{red}Numerical} 
  calculation of the relative contribution functions $C_i$ and $C_v$ of Stokes parameters $I$ and $V$ for the numerical solution of the H$\beta\lambda$4861.34\AA{} line for the VAL atmospheric model in the quiet Sun by \cite {Vernazza81} (\textbf{left}) and sunspot umbral atmospheric model by \cite{Ding91,Zhang20} (\textbf{right}) at wavelengths $\Delta\lambda$ = 0.45 (cross), 0.25 (star), 0.15 (diamond), 0.06 (delta), and 0.0  (black) \AA{} from the H$\beta$ line center. B = 1000 Gauss, $\psi$ = 30$^\circ$, azimuth $\varphi$ = 22.5$^\circ$, and $\mu =1$, while $\tau_c$ is the continuum optical depth at 5000 \AA{}. The horizontal coordinate is in the logarithmic scale. The calculated data resource is from~\cite{Zhang19,Zhang20}.
\label{fig:contribHbeta}}
\end{figure}
 
\subsection{Magnetic Field of Dark Filaments}

Solar dark filaments (prominences) consist of low-temperature and high-density plasma suspended above the chromosphere, which presents a dark shape against the relative bright background on the solar chromosphere in the solar disk. Although these seem low-key, they contain great energy. When the solar activity is intense, dark filaments (prominences) may erupt, and the released energy affects the whole solar system in the form of solar flares and coronal mass ejections, showing the unknown majestic power of the Sun. 

\begin{figure}[H]
\begin{minipage}[]{0.340\columnwidth}
{\includegraphics[width=0.95\columnwidth]{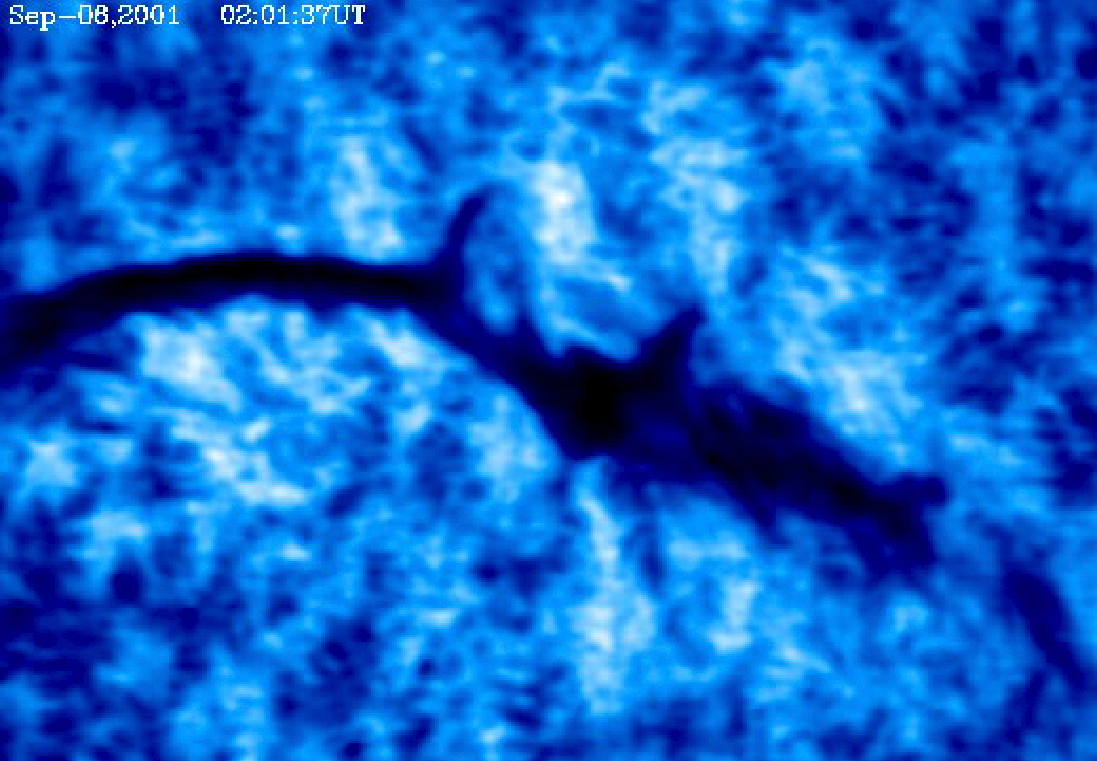}\vspace{0.1cm}}
{\includegraphics[width=0.95\columnwidth]{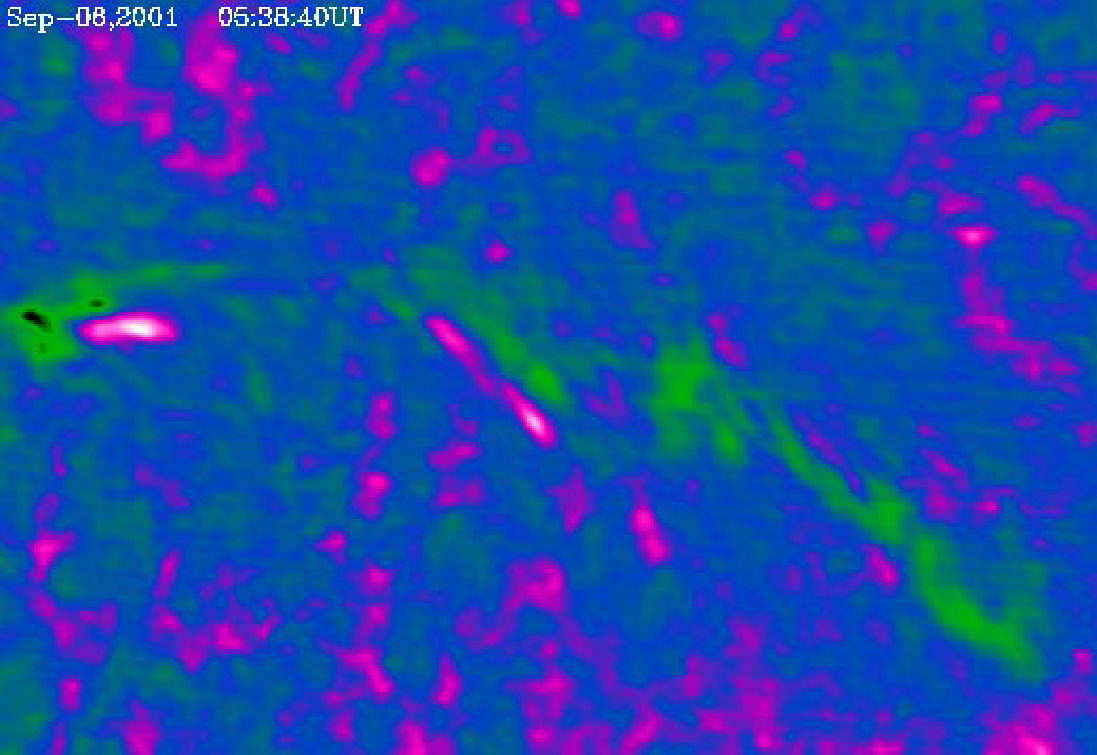}}
 \end{minipage}%
  \begin{minipage}[]{0.340\columnwidth}
{\includegraphics[width=0.95\columnwidth]{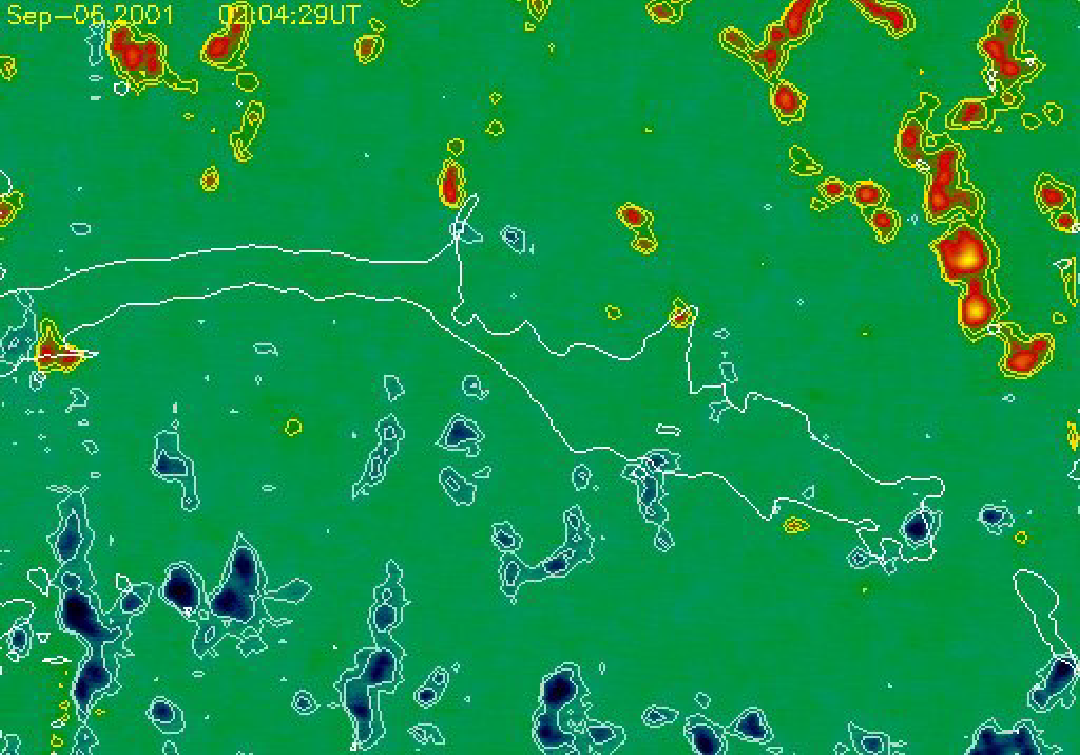}\vspace{0.1cm}}
{\includegraphics[width=0.95\columnwidth]{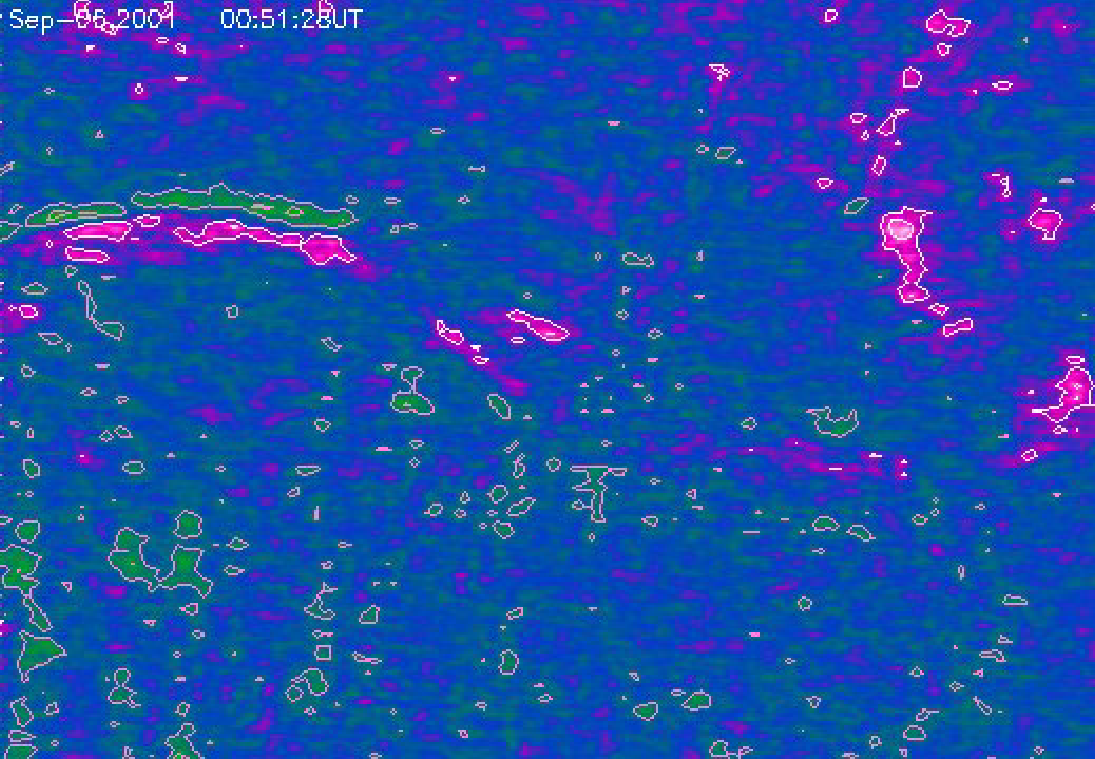}}
  \end{minipage}
  \caption{H$\beta$ 
  {filament} 
 and corresponding magnetograms and Dopplergram on 6 September 2001. (\textbf{Top left}):  H$\beta$ filtergram. (\textbf{Top right}): photospheric longitudinal magnetogram; the red (blue) contours correspond to positive (negative) fields of 100 and 200 Gauss. (\textbf{Bottom left}): H$\beta$ Dopplrgram; purple (green) is down (up). (\textbf{Bottom right}):  H$\beta$  longitudinal magnetogram; the purple (blue) contours correspond to positive (negative) fields of 40 and 80 Gauss.
 Provided by  \citet{BaoX03}.      
  }
\label{fig:Hbfilam}
\end{figure}

Here, we only introduce the results on the measurements of the dark filaments at Huairou Solar Observing Station. \citet{BaoX03} presented the line-of-sight magnetic field in the chromosphere and photosphere for a large quiescent filament on the solar disk on September 6, 2001 using the Solar Magnetic Field Telescope at Huairou Solar Observing Station (Figure \ref{fig:Hbfilam}).  The filament was located on the neutral line of the large-scale longitudinal magnetic field in the photosphere and chromosphere. 
     
\textls[25]{A notable question is how to construct believable spatial configurations of magnetic lines of the field around the filaments and support their formation in the higher solar atmosphere. Thus far, basic knowledge on the structure of the magnetic field of solar prominences is still almost entirely based on the inference of magnetohydrodynamic~equilibrium.
}

\subsection{Challenge for Measurements of Coronal Magnetic Fields}

It is normally believed that the reconnection of the magnetic field of flare-coronal mass ejections (Flare-CMEs) occurs in the high solar atmosphere, i.e., the chromosphere and corona. This conclusion is based on morphological and theoretical analysis \cite{Priest00,liu11}.  The solar corona is the outermost layer of the solar atmosphere, consisting of optically thin and highly ionized hot plasma. In visible light, the solar corona is much dimmer than the photosphere, at only one millionth as bright, making it difficult to observe in daylight \cite{ShaF23}. Diagnosis and structural analysis of the magnetic field in the upper solar atmosphere is of great value, and there have been a series of studies in this field. 

However, until now it has been difficult to obtain believable observational evidence indicating how and where the real reconnection process of the magnetic field occurs in the high solar atmosphere, even if magnetic reconnection is considered a reasonable topic \cite{Priest00}. If the real coronal magnetic field is to be accurately measured, it is vital to understand the reconnection of magnetic~fields.

The appropriate base for observing the corona is considered to be a high mountain site with thin air. The key to measuring the coronal magnetic field and diagnosing the faint polarized scattered light of the coronal emission lines containing the message of the field is to find the best available ground base and develop high-quality equipment \cite{Land04}.  It is believed that potential candidate sites for observing the coronal magnetic field can be found in Western China, such as at Daocheng in Sichuan Province, among others \cite{LiuY18}. 

The Daocheng site is located in the eastern area of the Tibetan plateau in China, and is considered to have excellent daytime conditions \cite{Song18} and a low level of sky brightness thanks to its high altitude of around 4700 m a.s.l. It is worth mentioning that a 10 cm coronagraph has been run successively for 10 years at the Lijiang Station (altitude 3200 m a.s.l.) using the coronal green line, which has helped to confirm the close relation between coronal brightness and coronal magnetic field intensity \cite{ZhangX22}. The white light (WL) corona was successfully observed in 2021 with a 50 mm balloon-borne coronagraph \cite{LiuY21,LinJ23}. However, no convincing polarization evidence was found from the WL linear polarizer system, which indicates that more experiments are needed for weak coronal signal detection in future work. Such work promises to be significantly moved forward with the recent operation of a new 25 cm ground-based coronagraph and efforts for the space-borne coronal magnetism exploration based on EUV 
 wavelengths 211 \AA 
 \cite{ZhangX22}.

\section{Discussion of Problems Involving the Optical Systems of Magnetic Field Observation Instruments}  

\subsection{Deviation in Observations and Measurements of Video Magnetographs}

A birefringence filter of $n$ levels optical elements is used at the Huairou magnetograph (Solar Magnetic Field Telescope) at Huairou Solar Observing Station of the National Astronomical Observatories in China. The transmission profile of the filter takes the following form~\citep{Ai82}:
\begin{equation}
\label{eq:filterHR}
\begin{aligned}
   T(\lambda)    =&  \prod ^{n}_{i=1} \cos^2\left(\pi\frac{\lambda-\lambda_0}{0.15\times2^{i-1}}+f_i\right)     
\end{aligned}\end{equation}
where $f_i$ is an adjustable parameter and $\lambda_0$ is the center wavelength of the selected transmission band. For the specific parameters and design of the solar magnetic field telescope, please refer to  \cite{Ai89,Zhang23}. 
When the Stokes spectrum passes through the filter, its amplitude is found to decrease due to the influence of the filter bandwidth. The amplitudes for different instruments and polarized spectral lines should change as well, that is to say, the magnitudes of Stokes parameters received by different magnetic field measuring instruments should be different.  

Generally, the crosstalk between the observed values of the four Stokes polarization parameters (I, Q, U, and V) and the real values of solar atmospheric radiation should be expressed as $\rm I'_{obs}=M \cdot I$ (Mueller matrix multiplied by Stokes vector) \citep{Stenflo94,Keller03}, where M is the Mueller matrix element of the $4\times 4$ matrix. This takes the form
\begin{equation}
\label{eq:Stoks}
S' (x, y, \lambda)=M_{ij} (x, y, \lambda) S(x, y, \lambda),    
\end{equation}
where $S' (x, y, \lambda)=(I', Q', U', V')_{obs} $, $S (x, y, \lambda)=(I, Q, U, V) $, and $M_{ij} (x, y, \lambda) $ are functions of the spatial {position (x, y)} and wavelength ($\lambda $).  $M_{ij} (x, y, \lambda)$ contains the distortion of the magnetic field information observed through imaging, which may have changes in position and amplitude when the Stokes parameter of radiation from the Sun passes through the~instrument.

The magnetic field signal modulation analyzers in solar magnetic field telescopes (magnetographs) are one of the core components for measuring the solar magnetic field~\mbox{\citep{aig81,Ai84}}, and often have effects on the accuracy of magnetic field measurements. Typical theoretical and experimental analyses indicate that the amplitude of the Stokes parameter $V$ reflecting the longitudinal magnetic field is often at least one order of magnitude higher than the parameters $Q$ and $U$ when characterizing the transverse magnetic field \citep{Zhang19}. This indicates that the crosstalk of different Stokes parameters caused by system errors in the magnetic field signal modulation analyzer will significantly deviate from the determination of the direction and magnitude of the transverse magnetic field \citep{Su07}. This issue in telescopes indicates the need for further analysis in addition to precise measurements and analysis bias when observing the solar magnetic field.

\subsection{Problems in Wide Field of View for Magnetic Field Observation}

When incident polarized light passes through an optical instrument, it is typical for the different Stokes parameters to interfere with each other. If we only consider the crosstalk in Stokes V measurement, the general expression from the fourth line {of the Stokes Equation~(\ref{eq:Stoks}) can be written approximately in the following form: 
\begin{equation}
V_{obs}\hspace{0.2em}\approx V+M_{41}{\cdot}I+M_{42}{\cdot}Q+M_{43}{\cdot}U\label{eq:wsz1}
\end{equation}
where $M_{44}\approx 1$ is set. For a detailed discussion of Equation~(\ref{eq:wsz1}), interested readers can refer to \cite{Wangx10}.
The usual case for the measurements of filter magnetograph is to normalize the observed Stokes $Q$, $U$, and $V$ signals using the intensity, i.e., $Q/I$, $U/I$, and $V/I$,} then spatially average them within the unresolved pixel or ``man-made defined'' area. 
{Thus, the units of the calibration coefficients and correction parameters are in Gauss.}

Figure \ref{fig:wszfig7} shows Huairou Solar Observing Station of the National Astronomical Observatories at different wavelength positions: $-$12 m\AA, +12 m\AA, $-$7 m\AA{} and $ +$7  m\AA. The observation times are also different, on December 2008 and June 2009. The original magnetic map is displayed in the upper panel of Figure \ref{fig:wszfig7}, and its corresponding corrected magnetic map is displayed in the lower panel, using one calibration template for the observed magnetograms. The correction effect in Figure \ref{fig:wszfig7} is enhanced because the grayscale of all magnetic maps is limited to $\pm$50 Gauss.

\begin{figure}[H]
\centering{\includegraphics[scale=0.75]{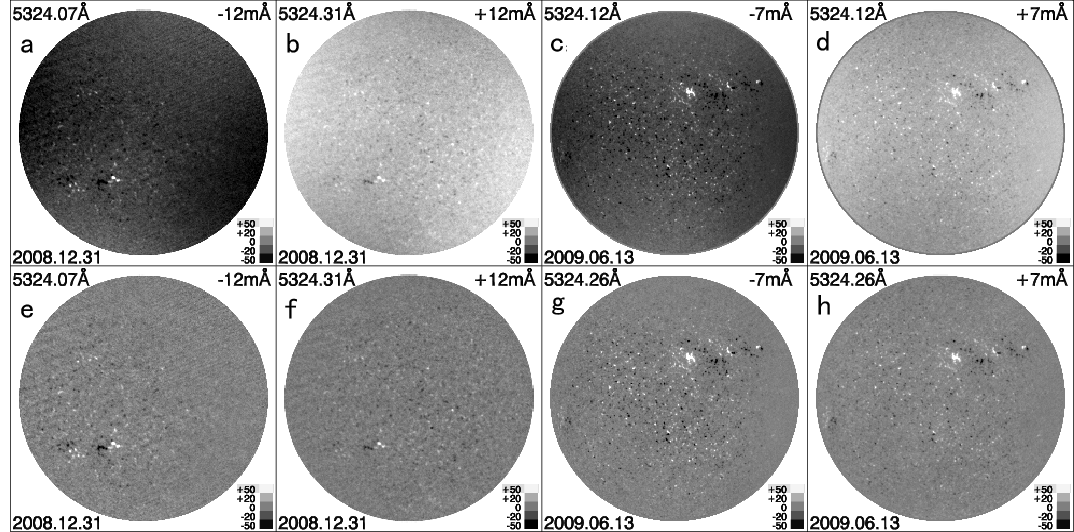}}
\caption{
{\color{red}The} 
 top panel shows the magnetograms before template correction, while the lower panel shows the corresponding magnetograms after template correction. The greyscale in the whole picture is limited to 50 Gauss in order to strengthen the display contrast. Note that the crosstalk correction template are the same. Due to different calibration signs, the blue wing magnetogram appears dark before correction, while the red one appears bright.  The solar-originated Doppler field affects the magnetic sensitivity and distribution of the calibration coefficients on the full-disc magnetogram of a single wavelength position. Taken from \cite{Wangx10}.}
\label{fig:wszfig7}

\end{figure}

Table \ref{tab:wsz01} lists some calibration coefficients at ten wavelength positions along with the minimum and maximum I--V crosstalk correction corresponding to these observing wavelength positions. From this table, it is clear that while $M_{41}$ is the same matrix for all wavelengths, the ``crosstalk magnetic field'' is different because of different calibration coefficients. Generally, the magnetic field error due to $M_{41}$ is a systematic error, and its amplitude is about 50 G.
Here, it can be seen that there are spatial inhomogeneities on the full disk magnetograms. This inhomogeneity comes from the difference in the Doppler shift of observation lines at different positions on the solar surface caused by the Sun's rotation, as well as from the difference in the off-axis properties of birefringence crystals at different positions of the filter due to the wide field of view effect.

\begin{table}[H]
\caption{\bf{The} 
 SMAT calibration coefficients of longitudinal magnetic field (units: Gauss), from 
\cite{Wangx10}. \label{tab:wsz01}} 

\setlength{\cellWidtha}{\textwidth/6-2\tabcolsep-0in}
\setlength{\cellWidthb}{\textwidth/6-2\tabcolsep-0in}
\setlength{\cellWidthc}{\textwidth/6-2\tabcolsep-0in}
\setlength{\cellWidthd}{\textwidth/6-2\tabcolsep-0in}
\setlength{\cellWidthe}{\textwidth/6-2\tabcolsep-0in}
\setlength{\cellWidthf}{\textwidth/6-2\tabcolsep-0in}
\scalebox{1}[1]{\begin{tabularx}{\textwidth}{>{\centering\arraybackslash}m{\cellWidtha}>{\centering\arraybackslash}m{\cellWidthb}>{\centering\arraybackslash}m{\cellWidthc}>{\centering\arraybackslash}m{\cellWidthd}>{\centering\arraybackslash}m{\cellWidthe}>{\centering\arraybackslash}m{\cellWidthf}}
\toprule
Blue wing &$-$0.15{\rm \AA} &$-$0.12{\rm \AA} &$-$0.10{\rm \AA} &$-$0.05{\rm \AA} &$-$0.02{\rm \AA}\\
Cali. coeff. &24,806 &19,728 &16,402 &14,316 &30,568 \\
$M_{41}$ correction &$-$15$\sim$$-$67 &$-$12$\sim$$-$53 &$-$10$\sim$$-$44 &$-$9$\sim$$-$39 &$-$19$\sim$$-$83\\
\midrule
Red wing &+0.02{\rm \AA} &+0.05{\rm \AA} &+0.10{\rm \AA} &+0.12{\rm \AA} &+0.15{\rm \AA}\\
Cali. coeff. &$-$26,764 &$-$12,386 &$-$12,249 &$-$16,393 &$-$21,197 \\
$M_{41}$ correction &16$\sim$73 &8$\sim$34 &7$\sim$33 &10$\sim$45 &13$\sim$58\\
\bottomrule
\end{tabularx}}
Cali. coeff. is the calibration coefficient of the full disk magnetogram. The $M_{41}$ correction is in units of Gauss. 
\end{table}

{

It is not possible to guarantee that the optical system of a full disk magnetic field telescope is completely telecentric in a realistic existing optical system.
The incident light from the different positions of the solar disk may be transmitted to slightly different positions of the optical system of the filter with slightly different inclined angles.} The variation of retardation caused by the spatial inhomogeneity of crystals themselves in the filter $\Delta \overline{n_{o-e}}+\delta n_{o-e}(x,y)=(n_o(x,y)-n_e(x,y))$ is also a source contributing to the errors of the signals in the magnetograms, where $\delta n_{o-e}(x,y)$ is caused by the spatial inhomogeneity. From this, we can infer the reasons for the corresponding changes in the sensitivity of polarized spectral lines at different positions in the plane in Figure \ref{fig:wszfig7}. This is due to the spatial inhomogeneity of the crystals being difficult to completely overcome in nature.

Moreover, we can discuss only the first order \citep{Evans40,Deng95,Stix02} with a thickness of $e$, which influences the narrow-band transmission profile of birefringent filters. The optical path difference between normal and extraordinary light with vertical incidence is \mbox{$\delta=2\pi e(n_o-n_e)/\lambda$}, where $\lambda$ is the vacuum wavelength. The difference $J=n_o-n_e$ is called the birefringence of the material. The wavelength of the maximum transmission is $\lambda_m=eJ/k$, where {$k$ is an arbitrary integer}. For large $k$, the distance from one maximum to the next is $\lambda\simeq eJ/k^2$.  Suppose that the paths of rays $\mathrm{Ia-Ia'}$ and $\mathrm{Ib-Ib'}$ passing through the filter crystal are different in terms of the filters or other birefringent optical systems, and that their optical path differences are different. Then, there will be a slight difference in the wavelength corresponding to the maximum transmission of the filter, which deviates from the original set wavelength in Formula (\ref{eq:filterHR}). Because the magnetic field in the solar disk is a surface source, this deviation of transmission wavelength is a distribution function of the surface space. When the average inclination angle of the beam $\mathrm{Ib-Ib'}$ at the thickest stage of the incident filter is $\theta$, the estimated magnitude of wavelength deviation of the transmission maximum is
\begin{equation}
\label{ }
\begin{aligned}
   \Delta\lambda_m&= \frac{e}{k}\left[\left(1+\frac{1}{2}\theta^2+\frac{5}{24}\theta^4\right)(n_o-n_e)_\theta -(n_o-n_e)_0\right]+\cdots\\
   &\approx\frac{e}{2k}\theta^2(n_o-n_e),     \qquad \mbox{as}   \qquad (n_o-n_e)_\theta -(n_o-n_e)_0\to 0,   
\end{aligned}
\end{equation}
where $(n_o-n_e)_\theta$ and $(n_o-n_e)_0$ represent the average refractive index difference of the oblique and vertical incident paths of ordinary and extraordinary light, respectively. The following provides a reference for the FeI$\lambda5324$ \AA{} wavelength used at Huairou Solar Observing Station of the National Astronomical Observatories. When the refractive index difference of birefringent crystal is $\Delta n=(n_o-n_e)=0.17421$, the filter corresponds to the longest crystal level thickness $ e = 70.804 mm$. If $ \theta = 0.5^\circ $, then the position with the maximum is shifted by $ \Delta \lambda = 0.2 $ \AA, while if $ \theta = 0.26^\circ $, then $\Delta\lambda=0.05$ \AA. Only the estimated value is provided here, and a more accurate analysis must still be carried out. For the Stokes spectrum of the FeI$\lambda5324$ \AA{} spectral line with an equivalent width of 0.334 \AA{} \citep{Ai82}, such errors are non-negligible.

When discussing the Stokes parameters $I$, $Q/I$, $U/I$, and $V/I$ emitted from the solar atmosphere, the errors in spatial uniformity distribution on the field of view and the refractive medium of the filter may lead to differences in the central wavelength $\lambda_0$ of the transmission band of the filter shown in Formula (\ref{eq:filterHR}) at different positions. For these reasons, the sensitivity of measurement signals in different areas of the panoramic magnetic map obtained under the wide field of view may differ. Thus, whether it is possible to ignore the off-axis effect of the filter and the influence of crystal inhomogeneity in the telecentric optical path design of the full solar-disk magnetic field telescope \citep{Zhang07} is a question worth discussing. Ways to remove the non-uniformities in observed Stokes signals in full-disk video magnetograms have been presented by  \citet{Su07} for different wavelengths of the magnetic sensitive line.  

Continuous observation at different wavelengths of magnetic sensitive spectral lines is carried out in Huairou Observing Station of the National Astronomical Observatories in order to make up for the influence of the uneven sensitivity of Stokes signals on a single wavelength in the wide field of view (full heliosphere) case. Stokes spectral analysis is of great value when seeking to accurately measure the magnetic field. The topic of wide field of view filters for measuring the solar magnetic field has been deeply studied at Huairou Observing Station.

\section{Further Discussion}  
Great progress has been made in the measurement of solar magnetic fields in China, and a series of research results have been obtained based on observations. When we analyze these carefully, we find that there is still room for further discussion around inaccuracies in measurement and analysis. These include the formation theory of magnetic sensitive lines in the solar atmosphere with magnetic fields as well as the development of technology for accurate analysis of the solar magnetic field. From the analysis provided in this paper, it is apparent that the differences between ground-based and space-based solar observation instruments are due to differences in observational conditions such as visibility at ground-based observation sites, which can often be seen in the measured results. These differences influence quantitative analysis of the magnetic field, for example when calculating current and magnetic helicity through the observed solar photospheric vector magnetic field.  

There are many topics worth discussing in the context of solar magnetic field measurement, for example how to solve the $180^\circ $ uncertainty of the photospheric transverse magnetic fields, which is a difficult problem caused by defects in the polarized light of the diagnostic magnetic field itself. Another question is how to deal with the consistency problem caused by the different noise levels of the photospheric longitudinal and transverse components of the measured magnetic field. 
Measurement of the solar chromospheric magnetic field and coronal magnetic field involves diagnosing scattering polarized light in dilute ionized gas and weak magnetic field  conditions, accurately analyzing and parameterizing the magnetic field in the nonlocal thermodynamic equilibrium state, and accurately diagnosing the changes in the magnetic field before and after solar flares, among other challenges. 

All of these topics are worthy of in-depth study from the aspects of both theory and observation, and are implied in the problems we discuss here in relation to the development of new magnetographs. In this case, it is still necessary to develop new systems for diagnosing magnetic fields in order to expand from single wavelengths to more in-depth observation and research on the Stokes polarized spectrum, including for video magnetographs\hl{.} 

\vspace{6pt}

\authorcontributions{
This research article encompasses the contributions of various authors. Hongqi Zhang, Jiangtao Su, and Mingyu Zhao have contributed to the analysis of the radiative transfer of spectral lines in the solar atmosphere and also worked on some of the vector magnetograms. Xingming Bao focused on the chromospheric magnetogram and explored its relationship with filaments. Yu Liu was responsible for the coronal part. Xiaofan Wang dealt with the full disk magnetograms, and Yingzi Sun made contributions regarding the magneograph.

}

\funding{
   
This study is supported by 
{grants} 
 from the National Natural Science Foundation (NNSF) of China under project grants 12073041, 12273059, and 12373063. 
}

\dataavailability{
}

\acknowledgments{The authors would like to thank the referees for their comments and suggestions as well as the staff at Huairou Solar Observing Station, the National Astronomical Observatories, Chinese Academy of Sciences. }

\conflictsofinterest{
The authors declare no conflicts of interest.

} 
\clearpage

\begin{adjustwidth}{-\extralength}{0cm}

\reftitle{References}

%

 

%


\PublishersNote{}
\end{adjustwidth}
\end{document}